\documentclass[twocolumn, nofootinbib]{revtex4-1}
\usepackage{graphicx,amsmath}
\usepackage{relsize}
\begin{document}
\title{The stability of transport models under changes of resonance parameters \\A UrQMD model study}
\author{Jochen Gerhard}
\email{jochen.gerhard@compeng.uni-frankfurt.de}
\author{Bj\o{}rn B\"auchle}
\author{Volker Lindenstruth}
\author{Marcus Bleicher}
\affiliation{Frankfurt Institute for Advanced Studies, Ruth-Moufang-Stra\ss{}e 1, 60438 Frankfurt am Main}
\affiliation{Institut f\"ur Theoretische Physik, Johann Wolfgang Goethe-Universit\"at, Max-von-Laue-Stra\ss{}e 1, 60438 Frankfurt am Main}
\affiliation{Institut f\"ur Informatik, Johann Wolfgang Goethe-Universit\"at, Robert-Mayer-Stra\ss{}e 11--15, 60054 Frankfurt am Main}

\begin{abstract} 
The Ultrarelativistic Quantum Molecular
Dynamics (UrQMD)
model is widely used to simulate heavy ion collisions in broad energy
ranges. It consists of various components to implement the different
physical processes underlying the transport approach. A major building
block are the shared tables of constants, implementing the baryon masses
and widths. Unfortunately, many of these input parameters are not well
known experimentally. In view of the upcoming physics program at the
Facility for Antiproton and Ion Research (FAIR),
it is therefore of fundamental interest to explore the stability of the
model results when these parameters are varied. We perform a systematic
variation of particle masses and widths within the limits proposed by
the particle data group (or up to $10\%$). We find that the model results
do only weakly depend on the variation of these input parameters. Thus,
we conclude that the present implementation is stable with respect to
the modification of not yet well specified particle parameters.
\end{abstract}
\maketitle
\section{Introduction}
One of the major themes of todays high energy physics is the exploration
of Quantum Chromo-Dynamics (QCD). QCD predicts that at sufficiently high
temperatures and densities, nuclear matter could exhibit a phase
transition into a new state of matter the Quark Gluon-Plasma (QGP). In
this state the usual color confinement is relaxed and the constituents
of the matter, namely quarks and gluons, are allowed to move over
distances larger than the scale of a single nucleon. Indeed, experiments
at the CERN-Super Proton Synchrotron (SPS), the BNL-Relativistic
Heavy Ion Collider (RHIC) and the Large Hadron Collider (LHC) at {CERN} have collected an
impressively large body of data that is consistent with the
interpretation that a QGP was formed for a short period of time. While
the current LHC program is running at the highest available energies
and therefore provides insights into the properties of the QGP at high
temperatures and very low baryon densities, the RHIC program is now
focused on the exploration of the QGP phase transition with a low energy
scan program. Here one hopes to find the existence and location of the
critical end point of the first order transition line that is expected
at high baryon densities. From the year 2018 on, this scan for the onset
of deconfinement and the search for the critical end point will also be
a top priority of the {FAIR} facility currently under construction in
Germany. 

Apart from the experimental difficulties, a major obstacle to pin down
the properties of strongly interacting matter is the unambiguous
interpretation of the experimental results. Unfortunately, first
principle lattice QCD calculations are currently only feasible in
thermal equilibrium and for very moderate $T/\mu_B$ values ($T$ being
the temperature and $\mu_B$ being the baryo-chemical potential).
Therefore, transport approaches like the Ultra-relativistic Quantum
Molecular Dynamics model (UrQMD), the Parton-Hadron-String-Dynamics
{(PHSD)} \cite{PHSB}, the Multi-Phase-Transport model {(AMPT)} \cite{AMPT}
and many other dynamical models are employed to link the final state
observables to the physics properties of the hot and dense stage of
the reaction. All these models have in common that they rely as input on 
measured quantities like the hadron masses, the hadron decay widths,
individual branching ratios and cross sections. Unfortunately, these
quantities are very often not exactly known, as one can see from an
inspection of the Particle Data Group (PDG) tables. 

In this paper we explore systematically how a variation of some of these
parameters influences the results obtained in transport simulation. As
an example, we employ the UrQMD model\cite{Bleicher,Bass,Petersen}. In the
long run, these investigations will allow to obtain a systematic error
of the simulations, which is needed to quantify the quality of the model
results. While we restrict ourselves to the UrQMD model in this study, the
results should (qualitatively) also be transferable to other transport
simulations based on similar physics assumptions as the ones mentioned
above.

\section{Computational set-up}
The idea is to replace the hard-coded hadron masses and widths with
automatically generated tables with varied parameters. Then a sufficient
number of simulations is performed and evaluated. To this aim, we have
designed \textsmaller{PYTHON} modules to read the up-to-date data from
PDG web page\cite{pdg}. The usage of \textsmaller{PYTHON} as \textit{glue code} for high
performance computations allows for very short development
cycles by combining rather slow, but powerful scripting parts of the
program with very fast compiled parts in another language. This
possibility is often used in graphics card computations\cite{casi}. 
We use \textsmaller{PYTHON} to automatically rewrite the \textsmaller{FORTRAN} source code of
UrQMD according to recalculated variations of the PDG data. This form
of meta programing is often preferable to using configuration
files. With the parameters being hard-coded in the \textsmaller{FORTRAN} source 
they are known at compile time, which enables the compiler to do more
optimizations.
We employ the Frankfurt {LOEWE-CSC}\cite{loewe} to carry out a
systematical parameter scan and to check of the stability of UrQMD. After
the computation the data files from different UrQMD runs are parsed
and compressed to statistics files, which can easily be interpreted on local systems. 
\begin{figure}[]
\includegraphics[width=0.5\textwidth]{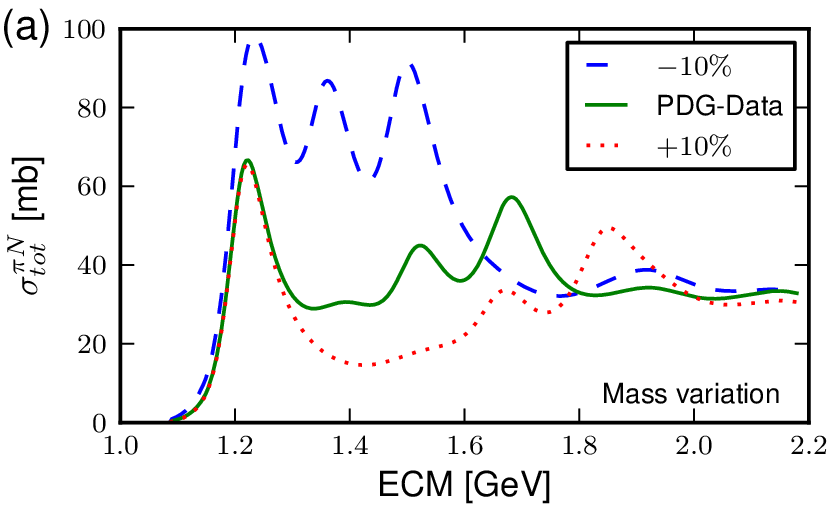}
\includegraphics[width=0.5\textwidth]{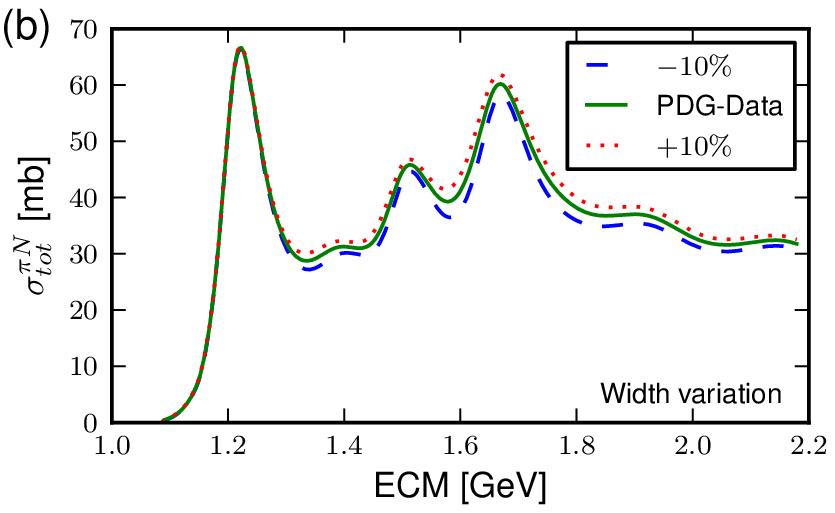}
\caption{\label{fig:sigtot_n}(Color online) (a): Total pion-nucleon cross section as a
function of $\sqrt s$ for a systematic variation of the nucleon
resonance masses.
The full line depicts the PDG averages, the dotted line shows a
variation of the PDG parameters by $+10\%$, the dashed line a variation by
$-10\%$. (b): Total pion-nucleon cross section as a function of $\sqrt
s$ for a systematic variation of the nucleon resonance widths. The full line
depicts the PDG averages, the dotted line shows a variation of the PDG
parameters by $+10\%$, the dashed line a variation by $-10\%$.}
\end{figure}

We explore the dependence of the UrQMD results on the particle data,
within the estimated errors provided by the PDG. In addition we vary the
parameters within an overall range of $\pm 10\%$ of mass and width.
Separate scans for variations of mass and width of each baryon family
are performed. Typically up to 10'000 events are simulated to stay clear of
statistical errors.  

\section{Cross sections}

Let us start by investigating the total pion-nucleon cross section as a
function of energy. This cross section is of special importance for the
dynamics of nuclear matter at intermediate energies. Resonances have
been investigated as probes for the interior of heavy ion reactions
\cite{Vogel}. It also serves as direct benchmark to adjust the
parameter sets since it is well measured experimentally. The total
cross section is given by
\begin{multline}
\sigma^{\rm tot}_{N\pi}= \sum_{R=\Delta,N^\ast} \langle j_N,m_N,j_\pi,m_\pi \|
J_R,M_R \rangle \\
\times \frac{2S_R+1}{(2S_N+1)(2S_\pi+1)} \frac{\pi}{p_{\rm CMS}^2}
\frac{\Gamma_{R\rightarrow N\pi}\Gamma_{\rm
tot}}{(M_R-\sqrt{s})^2+\frac{\Gamma_{\rm tot}^2}{4}}
\quad,
\end{multline}
with the total and partial decay widths $\Gamma_{\rm tot}$ and
$\Gamma_{R\rightarrow N\pi}$.
Thus, the cross-section depends on the widths and masses of all
nucleon- and Delta-resonances $N^\ast$ and $\Delta^{(\ast)}$.
Figure~\ref{fig:sigtot_n} (a) depicts the total pion-nucleon cross
section $\sigma_{\textnormal{tot}}^{\pi\textnormal{N}}$ as a
function of the center of mass energy $\sqrt s$ for a systematic variation of the nucleon
resonance masses. The
full line depicts the PDG averages, the dotted line shows a variation of the
PDG parameters by $+10\%$, the dashed line a variation by $-10\%$. One
clearly observes that the $\pi\textnormal{N}$ cross section varies
strongly if the resonance masses are changed. In turn, however, this
allows to pin down the resonance masses rather precisely. In contrast
a change of the resonance widths leaves the cross section unaltered.
Figure~\ref{fig:sigtot_n} (b) shows the total pion-nucleon cross
section as a
function of $\sqrt s$ for a systematic variation of the nucleon
resonance widths. The
full line depicts the PDG averages, the dotted line shows a variation of the
PDG parameters by $+10\%$, the dashed line a variation by $-10\%$.

\begin{figure}[]
\includegraphics[width=0.5\textwidth]{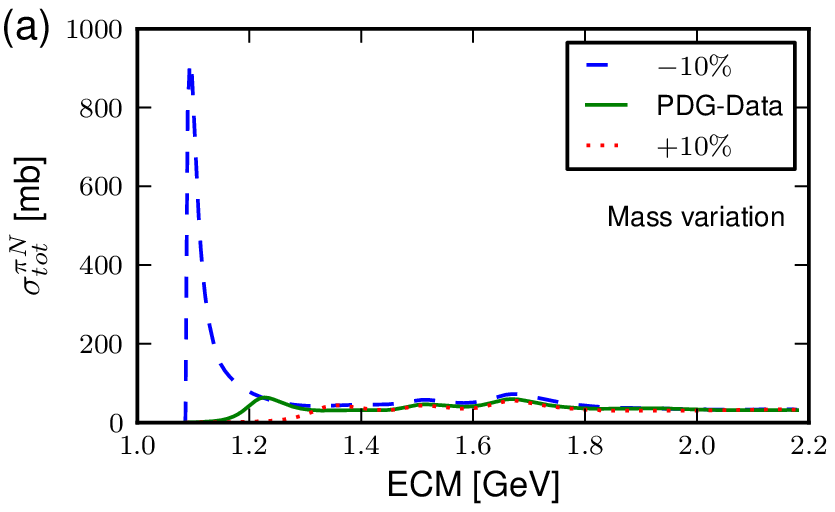}
\includegraphics[width=0.5\textwidth]{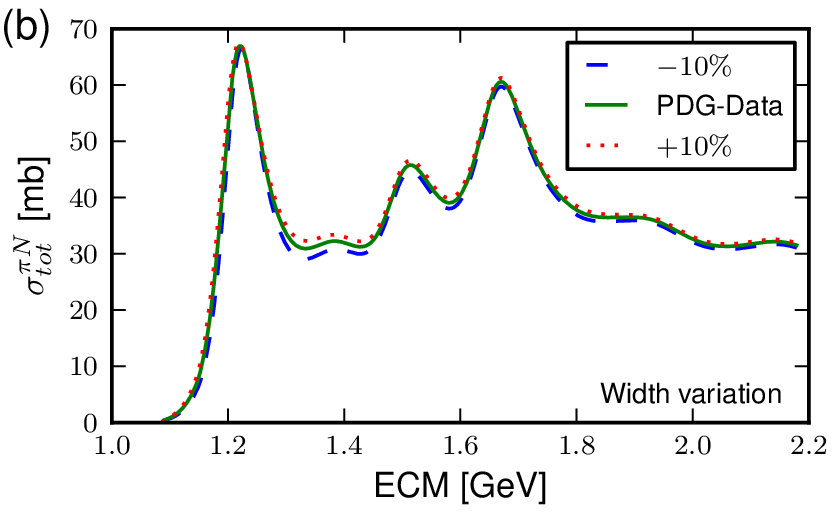}
\caption{\label{fig:sigtot_d}(Color online) (a): Total pion-nucleon cross section as a
function of $\sqrt s$ for a systematic variation of the $\Delta$ masses.
The full line depicts the PDG averages, the dotted line shows a
variation of the PDG parameters by $+10\%$, the dashed line a variation by
$-10\%$. (b): Total pion-nucleon cross section as a function of $\sqrt
s$ for a systematic variation of the $\Delta$ widths. The full line
depicts the PDG averages, the dotted line shows a variation of the PDG
parameters by $+10\%$, the dashed line a variation by $-10\%$.}
\end{figure}

Figure~\ref{fig:sigtot_d} (a) depicts the total pion-nucleon cross
section as a function of $\sqrt s$ for a systematic variation of the
$\Delta$ masses. The full line depicts the PDG averages, the dotted line
shows a variation of the PDG parameters by $+10\%$, the dashed line a
variation by $-10\%$. Figure~\ref{fig:sigtot_d} (b) shows the total
pion-nucleon cross section as a function of $\sqrt s$ for a systematic
variation of the $\Delta$ widths. The full line depicts the PDG
averages, the dotted line shows a variation of the PDG parameters by
$+10\%$, the dashed line a variation by $-10\%$.
One clearly observes that a variation of the nucleon masses has a
drastic effect on the pion-nucleon cross sections especially in the
$\Delta(1232)$ region. In comparison to the
available experimental data, strong constraints on the model parameters
can be obtained. In fact, the employed parameters are based on the PDG
data and re-adjusted within the limits of the PDG ranges. 

\section{Pion production}
\subsection{Pion yields}
\begin{figure}[]
\includegraphics[width=0.5\textwidth]{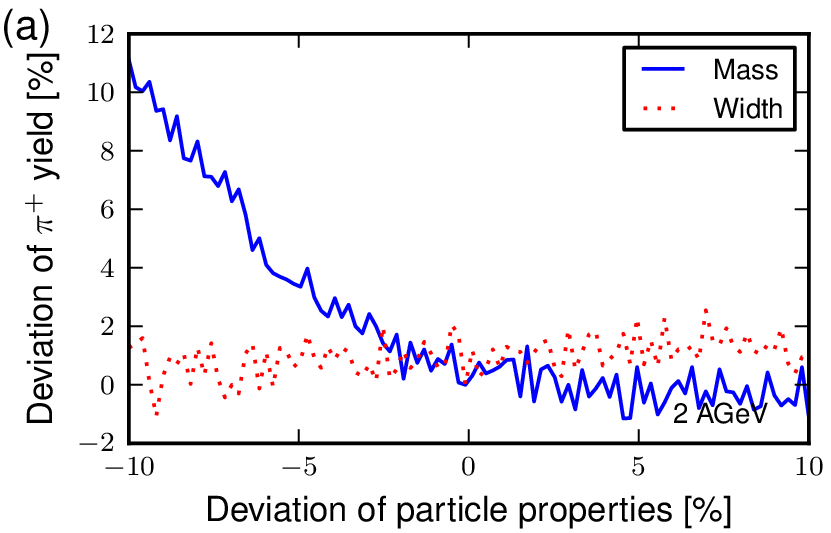}
\includegraphics[width=0.5\textwidth]{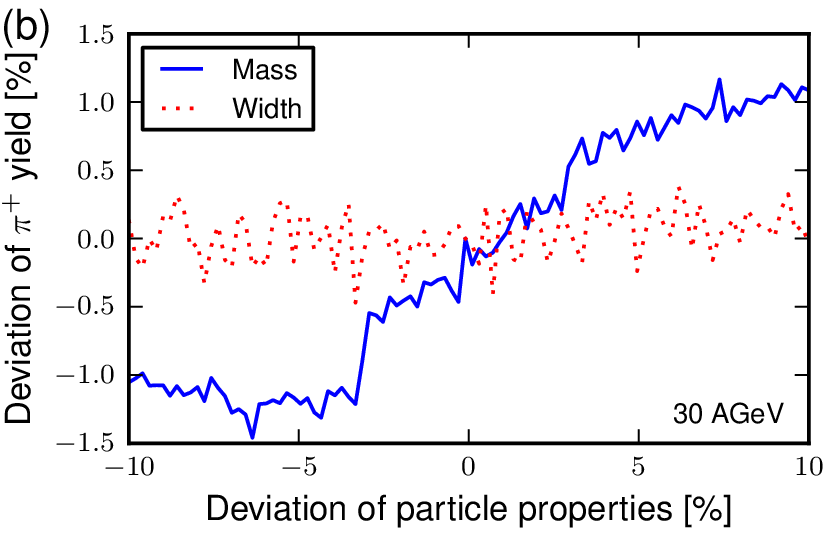}
\caption{\label{fig:pion_n}(Color online) (a): Relative pion yield in Pb+Pb collisions at 2A~GeV
beam energy. For a systematic variation of the masses and widths of the nucleon
resonances by $\pm 10\%$. (b): Relative pion yield in Pb+Pb collisions at 30A~GeV
beam energy. For a systematic variation of the masses and widths of the nucleon
resonances by $\pm 10\%$.}
\end{figure}
\begin{figure}[]
\includegraphics[width=0.5\textwidth]{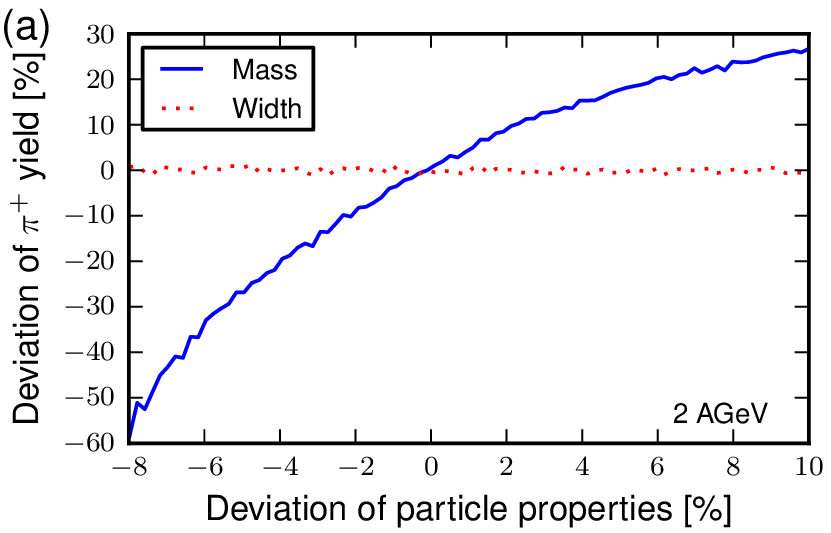}
\includegraphics[width=0.5\textwidth]{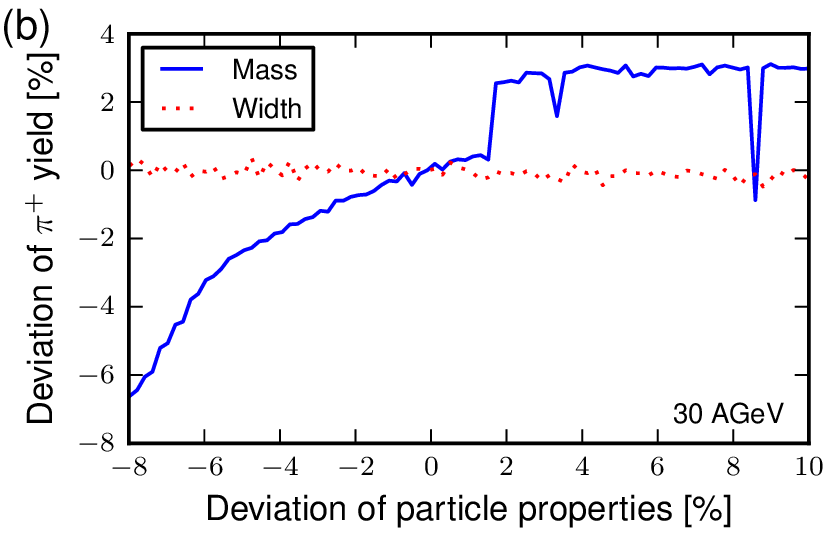}
\caption{\label{fig:pion_d}(Color online) (a): Relative pion yield in Pb+Pb collisions at 2A~GeV
beam energy. For a systematic variation of the masses and widths of the $\Delta$
resonances by $\pm 10\%$. (b): Relative pion yield in Pb+Pb collisions at 30A~GeV
beam energy. For a systematic variation of the masses and widths of the $\Delta$
resonances by $\pm 10\%$.}
\end{figure}

Let us next turn to the investigation of full Pb+Pb collisions and
focus on the {FAIR} energy range of 2 AGeV and 30 AGeV. Here we
investigate the total pion yield for a systematic variation of all
nucleon resonance masses $m_{N^\ast}$ by up to 10\%. We show the deviation of the pion yield
compared to a UrQMD calculation with the mean values of
the PDG data
files. Figure~\ref{fig:pion_n} (a) shows the
relative pion yield in Pb+Pb collisions at 2A~GeV beam energy for a systematic
variation of the masses and widths of the nucleon resonances by $\pm
10\%$. Figure~\ref{fig:pion_n} (b) shows the pion yield in Pb+Pb collisions at
30A~GeV beam energy for a systematic variation of the masses and widths of the
nucleon resonances by $\pm 10\%$.

Even at the lowest energy, which is strongly dominated by resonance
dynamics, the model results do at worst vary
linearly with the variation of the model parameters. At 30A~GeV, the
model results are stable against a variation of the resonance
parameters. The variation of the particle widths has no significant
effect on the model results.

Figure~\ref{fig:pion_d} investigates the pion production as a function of varying masses of the Delta-resonances
$m_\Delta$ and their widths $\Gamma_\Delta$. The masses of all Delta resonances have been scaled with the
same factor. Here we limit the variation to -8\% -- $+10\%$, because the
code becomes unstable for too low masses. Again a variation of the
width leaves the results unchanged. The variation of the
$\Delta^{(\ast)}$ masses, however results in a strong variation of the
pion yield at 2 AGeV. This effect is mainly attributed to the
$\Delta(1232)$ resonance that is pushed toward the kinematic limit
$(m_{\Delta(1232)}\rightarrow m_p + m_\pi)$. At 30 AGeV, the variance of
the yield stays generally moderate. However, the pion yield shows a
pronounced step if the masses are shifted by $\sim 1.8\%$ upward.
While the magnitude of the effect is small it indicates that complex
simulation models may exhibit discontinuous behaviors. Let us now have
a closer look into the origin of the step.  
\begin{figure}[]
\includegraphics[width=0.5\textwidth]{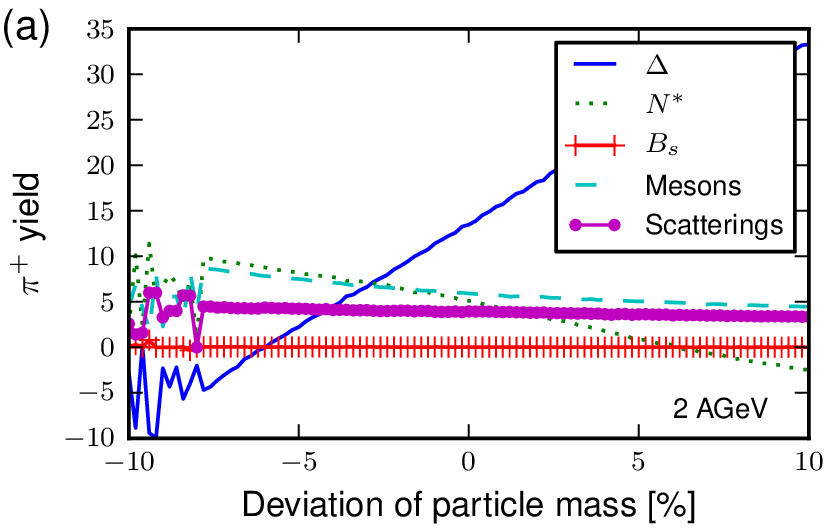}
\includegraphics[width=0.5\textwidth]{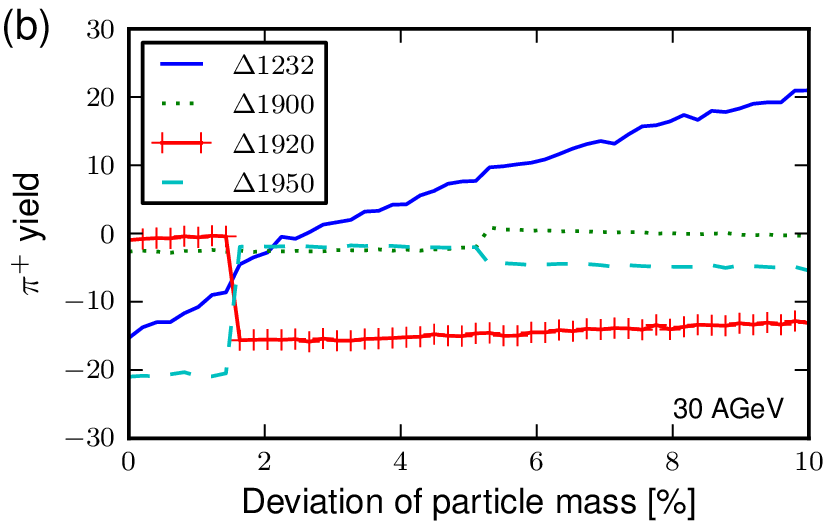}
\caption{\label{fig:channels}(Color online) (a): Pion yield in Pb+Pb collisions at 2A~GeV
beam energy itemizing different production processes. (b): Pion
production from various $\Delta$ resonances in Pb+Pb collisions at 30A~GeV.}
\end{figure}

For further analyses, we group different production processes into five classes, discriminating
the decay of Delta-resonances ($\Delta$), the
decay of nucleon-resonances ($N^\ast$),
the decay of strange baryons (accounting
for all unstable baryons not included in the former two classes)
($B_s$), the decay of meson resonances ($m$) and scatterings
($XY\rightarrow \pi+R$). In Figure~\ref{fig:channels} (a) one observes that
at $E_{\rm lab} = 2$A~GeV, the number of pions from scatterings
stays constant as a function of Delta mass, while the production of
pions from Delta decays rises linearly. At very low Delta masses (less
than
-6\% of the standard value), the formation of Deltas absorbs on
average more pions than are being produced by the decays thereof, while
at higher Delta masses, the opposite is true. The increase of pion
production from Deltas is counteracted by a decrease of pion production
from nucleon resonances $N^\ast$, which start to be net-absorbing above
+6\% of the standard (Delta-)masses. This can be explained in a picture of
detailed balance: When the Delta resonances produce more pions, the
equilibrium value of pion- and $N^\ast$-multiplicity is shifted toward
the $N^\ast$. Thus, the $N^\ast$-phase space is populated more quickly
than it is depleted. The same effect, though much weaker and not turning
around completely, can be seen in the decrease of the number of pions
from mesonic decays. In total, the rise of pions from Deltas counteracts
the fall of pions from $N^\ast$, thus leading to a weak overall rise of
the pion production.

At 30 AGeV we focus now on the step like behavior at a $\Delta$-mass
shift of $\sim +1.8\%$. Figure~\ref{fig:channels} (b) shows the contributions of different
$\Delta$-resonances to the final number of pions for varying Delta
masses between 0 and $+10\%$ at high impact energy $E_{\rm lab} =
30$A~GeV. In an analysis simlar to the one from
Figure~\ref{fig:channels} (a), we trace the step to
the Delta contribution. The step
we discovered earlier consists of an increased pion production (less absorption) from the
$\Delta_{1950}$-resonance, which rises from -20 to 0, and
a corresponding decreased production (increased net absorption) from the $\Delta_{1920}$-resonance,
which drops from 0 to -16. The
$\sqrt{s}$-distribution of the underlying $\pi N$ collision remains essentially unchanged as a
function of $m_\Delta$. Therefore, at the point of the discontinuities,
the $\Delta_{1920}$ takes the role that $\Delta_{1950}$ had
at lower masses. Since the branching ratios $\Delta_{1920} \rightarrow
\pi\Delta_{1232}$ and $\Delta_{1950} \rightarrow \pi\Delta_{1232}$
differ by a factor of 2 (40\% vs.\ 20\%), the number of $\Delta_{1232}$
and thus the number of pions are changed over a small mass intervall.  We find a
simliar behavior at $\sim +5\%$, where the pion production from
$\Delta_{1950}$ decreases, while the production from $\Delta_{1900}$ increases. 
Superimposed is an approximately
linear rise of pion production from the lowest Delta resonance
$\Delta_{1232}$. 

\subsection{$p_T$-spectra of $\pi^+$}
\begin{figure}[t!]
\includegraphics[width=0.5\textwidth]{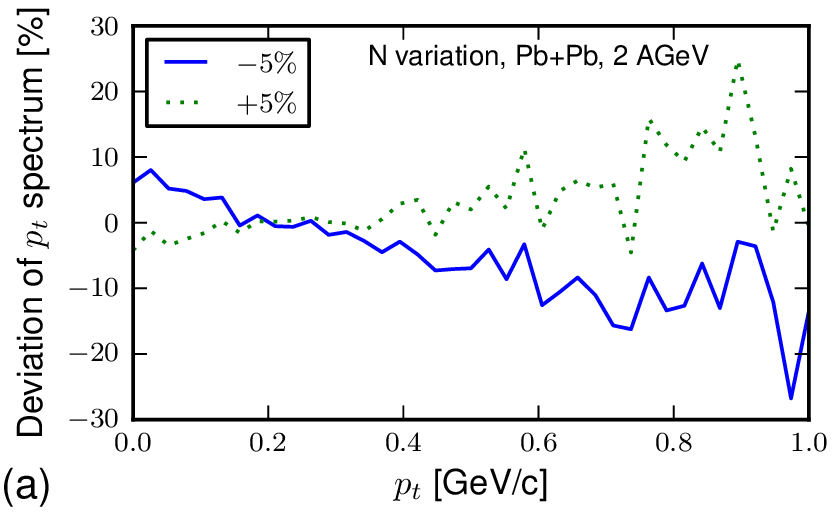}
\includegraphics[width=0.5\textwidth]{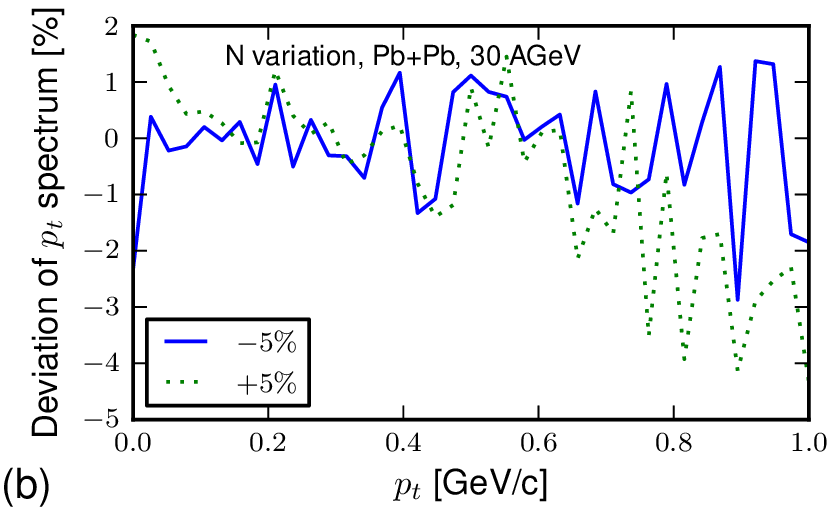}
\caption{\label{fig:pt_n}(Color online) (a): Deviation of pion transverse momentum spectra
  ($p_t$) in Pb+Pb collisions at 2A~GeV for variations of the nucleon resonance
 masses. (b): Deviation of pion transverse momentum spectra ($p_t$) in Pb+Pb
collisions at 30A~GeV for variations of the nucleon resonance masses.}
\end{figure}

Finally, we discuss the transverse momentum distributions\footnote{We
  also analyzed the rapidity spectrum distributions, but found no
  significant deviation.}. Here we
investigate the pions transverse momentum distributions in Pb+Pb collisions
at 2 and 30 AGeV.  Figure~\ref{fig:pt_n} (a) shows the deviation of the pion transverse
momentum spectra ($p_t$) in Pb+Pb collisions at 2A~GeV for variations of
the nucleon resonance masses.

At 2 AGeV, a decrease of the nucleon
resonance masses shifts the pions to lower $p_t$, while an increase of
the masses shifts it to higher transverse momenta. However, variations
of the yields at higher $p_t$ maybe up to $\pm 20\%$ for a variation
of $\pm 5\%$.

\begin{figure}[t]
\includegraphics[width=0.5\textwidth]{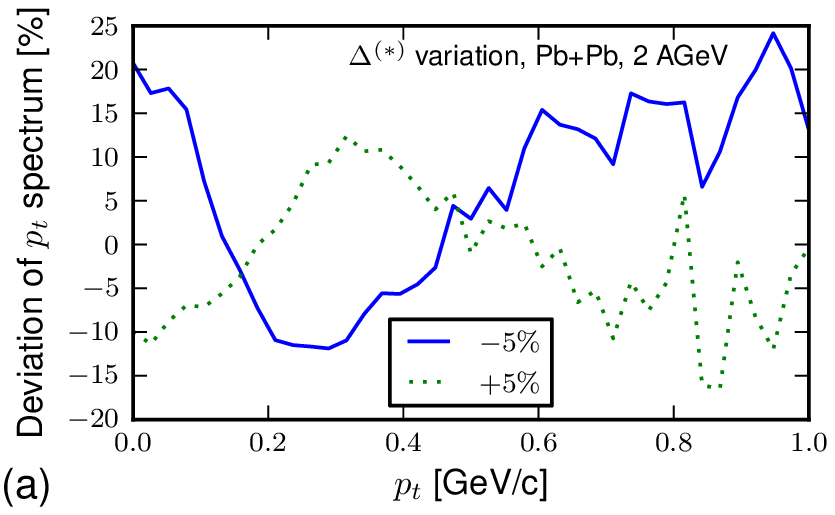}
\includegraphics[width=0.5\textwidth]{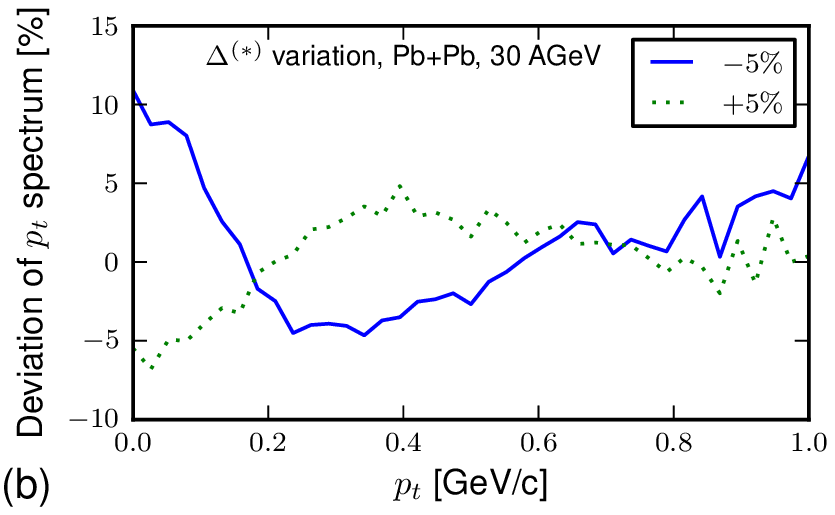}
\caption{\label{fig:pt_d}(Color online) (a): Deviation of pion transverse momentum spectra ($p_t$)
in Pb+Pb collisions at 2A~GeV for variations of the $\Delta$ resonance
masses. (b): Deviation of pion transverse momentum spectra ($p_t$) in Pb+Pb
collisions at 30A~GeV for variations of the $\Delta$ resonance masses.}
\end{figure}

Figure~\ref{fig:pt_n} (b) displays the
deviation of the pion transverse momentum spectra ($p_t$) in Pb+Pb collisions at 30A~GeV for
variations of the nucleon resonance masses.  Figure~\ref{fig:pt_d} (a)
shows the deviation of the pion transverse momentum spectra ($p_t$) in Pb+Pb collisions at
2A~GeV for variations of the $\Delta$ resonance masses.
Again, we do not observe and effect on the calculations at $E_{\rm lab} = 30$A~GeV is
within the statistical fluctuations.
Figure~\ref{fig:pt_d} (b) displays the deviation of the pion transverse momentum
spectra ($p_t$) in Pb+Pb collisions at 30A~GeV for variations of the $\Delta$
resonance masses.

The effect of variations in Delta resonance masses is strongly
non-~linear. Both at high beam energies
and at low beam energies, we can distinguish three transverse momentum
regions. Pions from intermediate transverse momentum $0.2 < p_t <
0.6$~GeV are being shifted to low transverse momentum $p_t < 0.2$~GeV,
if the masses are decreased and vice versa, if the masses are increased.
This is expected, since the available kinetic energy in a Delta decay
decreases with decreasing Delta mass.
At higher transverse momenta, lower masses lead to higher pion yields,
while higher masses lead to lower pion yields, which is a reversal from
the behavior observed from varying the nucleon resonance masses.
Furthermore, we observe the effects to be a lot stronger in low-energy
collisions. Also the variation of the $\Delta$ masses by $\pm 5\%$
results in modifications of the pion yield by $\pm 20\%$ in given
$p_t$ regions.

\section{Summary}
In light of the upcoming high precision experiments at {FAIR}, it is
highly desirable to obtain better estimates on the systematic errors
of transport simulations. We addressed this
question by using the UrQMD transport approach in nucleus-nucleus
reactions in the {FAIR} energy regime from 2 to 30A~GeV. We have analysed
elementary cross sections in pion-nucleon reactions, as well as lead-lead
collisions for various sets of input parameter variations of the hadron
masses and widths. Although the analyzed
quantities show globally only a weak dependence, discontinuities
like in Figure~\ref{fig:pion_d} (b) may occur and influence
predictions made by the applied models. The dependence is strongest in
low-energy collisions, where the collision dynamics is dominated by
resonance production and decay. Here, one may encounter systematic
errors on the order of $\pm 20\%$. At higher energies, the systematic 
errors are much smaller. One should note, that the present study
explored a \textit{worst case scenario} where all parameters were
shifted simultaneously in one direction. The error on the masses (and
widths) are however uncorrelated and should therefore induce smaller
systematic bias into the simulations as compared to this
study. Therefore, we conclude that the
predictive and analysis power of the present approach is better than a
systematic error of 20 \%.
\section{Acknowledgements}
The computational resources were provided by the {LOEWE-CSC}. This work
was supported by {HIC for FAIR} within the Hessian {LOEWE} initiative.
 
\bibliography{stability}
\end{document}